\documentclass[aps,prb,twocolumn,floatfix]{revtex4-1}

\usepackage{amsfonts}
\usepackage{amsmath}
\usepackage{bm}
\usepackage{amssymb}
\usepackage{subfigure}
\usepackage{graphicx}

\begin{document}

\title{Spin and Charge Fluctuations in the $\alpha $-structure Layered
Nitride Superconductors}
\author{Quan Yin}
\email{Present address: Department of Physics and Astronomy, Rutgers University, Piscataway, NJ 08854}
\author{Erik R. Ylvisaker}
\author{Warren E. Pickett}
\affiliation{Department of Physics, University of California, Davis, CA 95616}
\date{\today }

\begin{abstract}
To explore conditions underlying the superconductivity in electron-doped TiNCl
where T$_c$ = 16 K,
we calculate the electronic structure, Wannier functions and spin and charge 
susceptibilities using first-principles density functional theory.
TiNCl is the first high-temperature superconductor discovered in the $%
\alpha $-structure of the layered transition-metal nitride family MNCl
(M=Ti, Zr, Hf).
We construct a tight-binding model based on Wannier functions derived
from the band structure, and consider explicit electronic interactions
in a multi-band Hubbard Hamiltonian, where the interactions
are treated with the random phase approximation (RPA) to calculate spin and
charge susceptibility. The results show that, consistent with TiNCl being a nonmagnetic
material, spin fluctuations do not dominate over charge fluctuations
and both may have comparable impact on the properties of the doped system. 
\end{abstract}

\maketitle

\section{Introduction and Background}

High-temperature superconductivity has been a widely pursued subject for
condensed matter physicists for over two decades. There are several classes
of materials where unconventional superconductivity is found: cuprates whose
highest T$_{c}$'s remain unrivaled, the recently discovered and intensively
studied iron pnictides, Na$_{1-x}$CoO$_{2}$ intercalated with H$_{2}$O, BaBiO%
$_{3}$ doped by K, the Pu-based ``115'' heavy fermion series, and the
transition metal nitride halide\ MNX (M=Zr, Hf; X=Cl, Br, I). MNX
crystallizes in two structures, labeled $\alpha$ and $\beta$. 
The Zr and Hf members were found to be
superconducting with unprecedented critical temperatures for nitrides (15K,
25K) in the $\beta $-structure, which is isostructural to SmSI, and contains
double-honeycomb layers of alternating M and N atoms \cite{beta-structure}.
The sister compound TiNCl with $\alpha$-structure has now been discovered to
superconduct (16K) as well.\cite{TiNCl-SC}

Based on many examples now, layered structures seem to favor
high-temperature superconductivity. The reduced dimensionality promotes
various instabilities associated with Fermi surface nesting. Due to
electron-electron interactions, many parent compounds of HTSCs exhibit
long-range magnetic order, such as antiferromagnetism (AFM) in cuprates and
iron pnictides. The AFM order needs to be destroyed upon doping, by
electrons or by holes, to open the way for superconductivity. This is not
the case for non-magnetic MNX, which are band insulators with a band gap of $2-4$ eV, and
the transition metal $d$ states make up most of the lower conduction bands 
\cite{HfNCl-band}.

Both the $\alpha$- and $\beta$-polymorphs are quasi-2D structures 
with large interlayer spacing and weak van der Waals coupling between
layers. When doped with electrons, A$_{x}$MNX (A being alkali metals) remains insulating at low
concentration, then suddenly become superconducting at about $x=0.13$ in
HfNCl and $x=0.06$ in ZrNCl, and maintain a relatively constant T$_{c}$ up
to $x=0.5$ [\onlinecite{LixZrNCl-Tcdoping}]. The superconducting transition
temperature can be as high as $26$ K, discovered in Li$_{x}$(THF)$_{y}$HfNCl 
\cite{HfNCl-Nature1998}. It has also been found that T$_{c}$ may be
correlated with the interlayer spacing, which can be tuned by intercalation
of different sized molecules \cite{HfNCl-Tcspacing}. Also, the electron
doping can be substituted by ion vacancy of the Cl atoms with similar
superconductivity being found \cite{MNCl1-x-SC}. In a theoretical treatment
by Bill \textit{et al.}, the dynamical screening of electronic interactions
in these materials was modeled\cite{plasmon1,plasmon2} by conducting sheets spaced by dielectrics.\cite{Jain1985}
A fully open large superconducting gap without nodes
was observed with tunneling spectroscopy. \cite{LiZrNCl-SCgap,LiTHFHfNCl-gap1,LiTHFHfNCl-gap2}

Experimental observations from several perspectives confirm that MNX are not
electron-phonon BCS superconductors: (1) measured isotope effects are small;%
\cite{HfNCl-isotope,LixZrNCl-isotope} (2) specific heat measurements\cite%
{LixZrNCl-spheat} indicate a small mass enhancement factor; (3) the density
of states at the Fermi level is small (when electron doped), and T$_{c}$ is
almost independent of the doping level in the range $0.15<x<0.5$. As a
feature peculiar to this system, T$_c$ actually increases as the
metal-insulator transition at $x_{cr}$=0.06 is approached, rather than
following the common dome shape with doping. Linear response calculations
also agree on the small electron-phonon coupling constant \cite{LixZrNCl-phonon}
that cannot account for the observed T$_{c}$. The impressive high
transition temperatures and easy tunability of carrier concentrations (and
sometimes effective dimensionality) suggest there is potential to
reach higher T$_{c}$ in this class.\cite{Yamanaka-review}

The possible candidates of pairing mechanism responsible for the observed
high T$_{c}$ are spin and charge fluctuations, which have been discussed by
both experimentalists and theorists, but opinions remain controversial. The
specific heat measurement on Li$_{x}$ZrNCl is suggestive of relatively
strong coupling superconductivity, based on the observed large gap ratio and
specific jump\cite{LixZrNCl-spheat} at T$_c$. The inter-layer spacing
dependence of T$_{c}$ reveals the close relation between the pairing
interaction and topology of the Fermi surface \cite{HfNCl-Tcspacing},
implying a spin and/or charge induced superconductivity. The magnetic
susceptibility measurement of heavily doped Li$_{x}$(THF)$_{y}$HfNCl
indicates low carrier density and negligible mass enhancement factor, in
favor of charge fluctuations over the spin fluctuations.\cite%
{Li(THF)HfNCl-X-expt}

A detailed measurement of the doping dependence of specific heat
and magnetic susceptibility has been performed on Li$_{x}$ZrNCl and the data
were compared with calculations based on a model Hamiltonian \cite%
{LixZrNCl-Xs-expt,Kuroki-FLEX}, which shows correlation between T$_{c}$ and
magnetic susceptibility. Some change is occurring that affects both, but a
causal relationship has not been established. There have been several band
structure calculations for ZrNCl and HfNCl that provide the basis for more
specific studies, mostly in the superconducting 
$\beta$-structure.\cite{HfNCl-band,ZrHfNCl-band1,ZrHfNCl-band2,ZrNCl-band}

Recently, Yamanaka \textit{et al} \cite{TiNCl-SC} reported superconductivity
in the alkali metal intercalated $\alpha $-TiNCl, with T$_{c}$ up to $16.5$ K.
This is the first MNX compound found to be superconducting in the $\alpha $%
-structure. The lack of Fermi surface nesting in $\alpha $-Li$_{x}$TiNCl
seems to exclude large magnetic fluctuations, and its low carrier density
character is indicative of charge fluctuations as a more likely driving
force. Theoretically, charge fluctuation induced superconductivity has been
discussed in the Hubbard model \cite{ChargeFluc1,ChargeFluc2}, the $d$-$p$
model \cite{ChargeFluc3}, and been applied to study Na$_{x}$CoO$_{2}\cdot y$H%
$_{2}$O \cite{NaCoO2-FLEX,NaCoO2-RPA} and organic molecular superconductors 
\cite{ChargeFluc-organicSC}. A more detailed investigation of the electronic
structure and dynamical spin/charge susceptibility is needed to study the
superconducting mechanism in $\alpha $-TiNCl. In the present work, we
conduct a theoretical calculation of spin and charge susceptibilities using
a many-body Hamiltonian, based on a realistic band structure calculated by density
functional theory.

\section{Crystal Structure}

\begin{figure}[tbh]
\centering
\includegraphics[width=0.8\columnwidth]{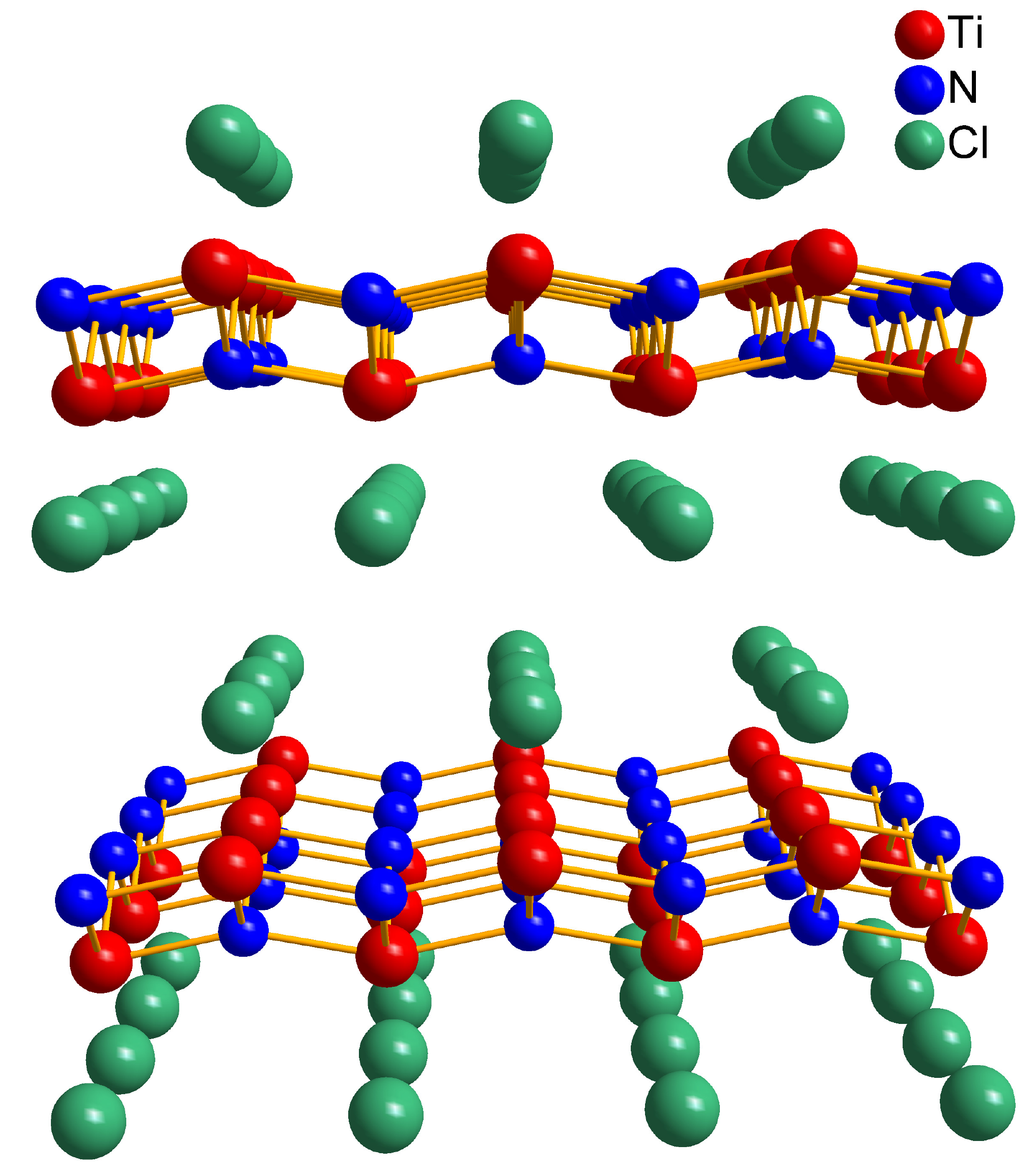}
\caption{Crystal structure of $\protect\alpha $ phase TiNCl (Pmmn, \#59).
Viewpoint is along the b-axis, perpendicular to the buckling of the Ti-N
net. The buckled Ti-N layer leaves each Ti coordinated (roughly
octahedrally) by 4 N and 2 Cl atoms. The Cl coordinates with Ti along the
b-axis, which accounts for its orthorhombic structure.}
\label{structure}
\end{figure}

The $\alpha$-structure\cite{TiNCl-SC} of the MNX class of compounds, often
called the FeOCl structure, is shown in Fig. \ref{structure}, with
structural data given in Table \ref{lattice}. The $\alpha$-structure TiNCl
belongs to space group \textit{Pmmn} (\#59), with 6 atoms per unit cell
occupying the following sites: Ti($2b$) (0,$\frac{1}{2},z_{Ti}$), N($2a$) ($%
\frac{1}{2},\frac{1}{2},z_{N})$ and Cl($2a$) (0,0,$z_{Cl}$). The generators
of Pmmn are two simple reflections $x\rightarrow -x$ and $y\rightarrow -y$,
and the non-symmorphic reflection $z\rightarrow -z$ followed by a ($\frac{1}{%
2},\frac{1}{2},0)$ translation.

The Ti-N net within TiNCl is topologically equivalent to that of a single
NaCl layer. There is strong buckling this Ti-N net perpendicular to the $b$
direction, such that neighboring chains which are directed along $a$ differ
in height. These chains are themselves somewhat buckled, all of this leading
to the placement of Ti ions $\pm$0.8 \AA ~from the average height, and N
ions $\pm$0.4 \AA ~from the average height. The Ti ions are two-fold
coordinated by Cl ions lying in the $y-z$ plane; the breaking of square
symmetry of the TiN layer by its strong buckling, can be regarded as ``due
to'' this positioning of the Cl ions.

Finally, each Ti is six-fold coordinated by four N and two Cl atoms. The two
Ti-N bonds have very close lengths of $2.008$ \AA ~and $2.015$ \AA ,
respectively, though the N ions lie at different heights in the $x$ and $y$
directions. Very roughly, the Ti ion is in octahedral coordination (see Fig.
5 of Ref. [\onlinecite{TiNCl-SC}]), with approximate axes (1,0,0)
(toward two neighboring N ions), and (0,1,1) and (0,1,-1) (each toward one N
and one Cl ion), and indeed a rough $t_{2g}-e_{g}$ splitting of the Ti $3d$
states results. The Ti-Ti distance is $3.003$~\AA , not much larger than
that of Ti-N due to the buckled layer structure, so in the tight-binding
model we construct in the next section, the hoppings between Ti sites are also
important.

The experimental lattice constants and atomic positions, and relaxed
structure parameters with respect to total energy, which are used in our
calculation, are listed in Table \ref{lattice}. The calculations (see below)
confirm the expected formal valences. The calculated lattice constants are
1-1.5\% smaller than the experimental values, but this has little effect on
the electronic structure. TiNCl is still calculated to be an ionic insulator
and its theoretical gap is similar to what is calculated using the
experimental lattice parameters.

\begin{table}[tbh] \centering%
\begin{tabular}{|l|c|c|c|c|c|c|}
\hline
& $a$ & $b$ & $c$ & $z_{Ti}$ & $z_{N}$ & $z_{Cl}$ \\ \hline
Expt. & \multicolumn{1}{|l|}{$3.938$} & \multicolumn{1}{|l|}{$3.258$} & 
\multicolumn{1}{|l|}{$7.800$} & \multicolumn{1}{|l|}{$0.1011$} & 
\multicolumn{1}{|l|}{$0.0509$} & $0.3322$ \\ 
Theory & \multicolumn{1}{|l|}{$3.891$} & \multicolumn{1}{|l|}{$3.200$} & 
\multicolumn{1}{|l|}{$7.699$} & \multicolumn{1}{|l|}{$0.1011$} & 
\multicolumn{1}{|l|}{$0.0522$} & $0.3384$ \\ \hline
\end{tabular}%
\caption{Lattice constants (in units of \AA) and internal structural parameter $z$ 
for the three atoms.
Experimental values are from Ref. \onlinecite{TiNCl-SC}.  The theoretical values
are our optimized values.}\label{lattice}%
\end{table}%

\section{Band Structure and Wannier Functions}

\subsection{Methods}

The band structure has been computed by the full-potential local orbital
minimal basis set method implemented in the FPLO code.\cite{FPLO} The
exchange correlation is treated by the generalized gradient approximation
GGA96,\cite{GGA96} and the \textbf{k}-mesh used is $16\times 16\times 8$. 
The effect of spin-orbit coupling is small so calculations were done
in the scalar relativistic scheme.

\begin{figure}[tbh]
\centering
\includegraphics[width=0.95\columnwidth]{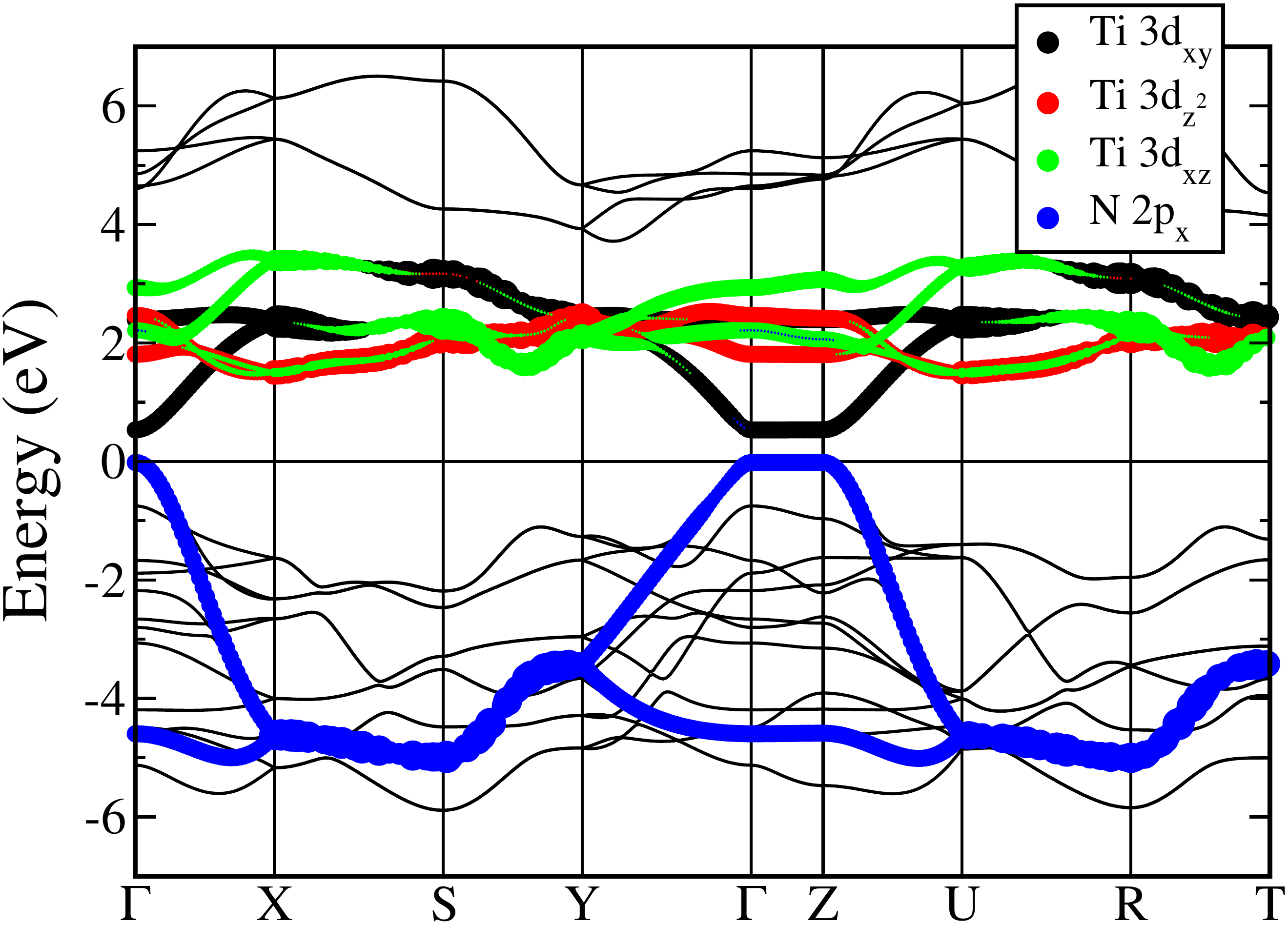}
\caption{(color online) Band structure of TiNCl along the orthorhombic
symmetry lines, calculated with the GGA exchange-correlation functional. The
thick ``fatbands'' are the tight-binding representation determined by the
Wannier functions labeled in the inset, chosen to represent accurately the
bands at and around the Fermi level after doping.}
\label{fig:band}
\end{figure}

\subsection{Electronic Structure}

The calculated band structure of pristine TiNCl is shown in Fig. 
\ref{fig:band} and is generally consistent with that presented by Yamanaka 
\textit{et al.}\cite{TiNCl-SC} plotted along other lines in the zone. It is
an insulator with a calculated energy gap of $0.5$ eV. The real band gap
may be as large as $1$ eV, based on the common observation that LDA and
GGA underestimates gaps in insulators. The band structure exhibits clearly a
two-dimensional feature, gauged from the general flatness of bands along 
the $\Gamma-Z$ direction perpendicular to the layers. The states on either 
side of the gap are very two-dimensional, considering the extreme flatness
of those bands along $\Gamma$-Z.

\begin{figure}[t]
\centering
\includegraphics[angle=-90,width=3.25in]{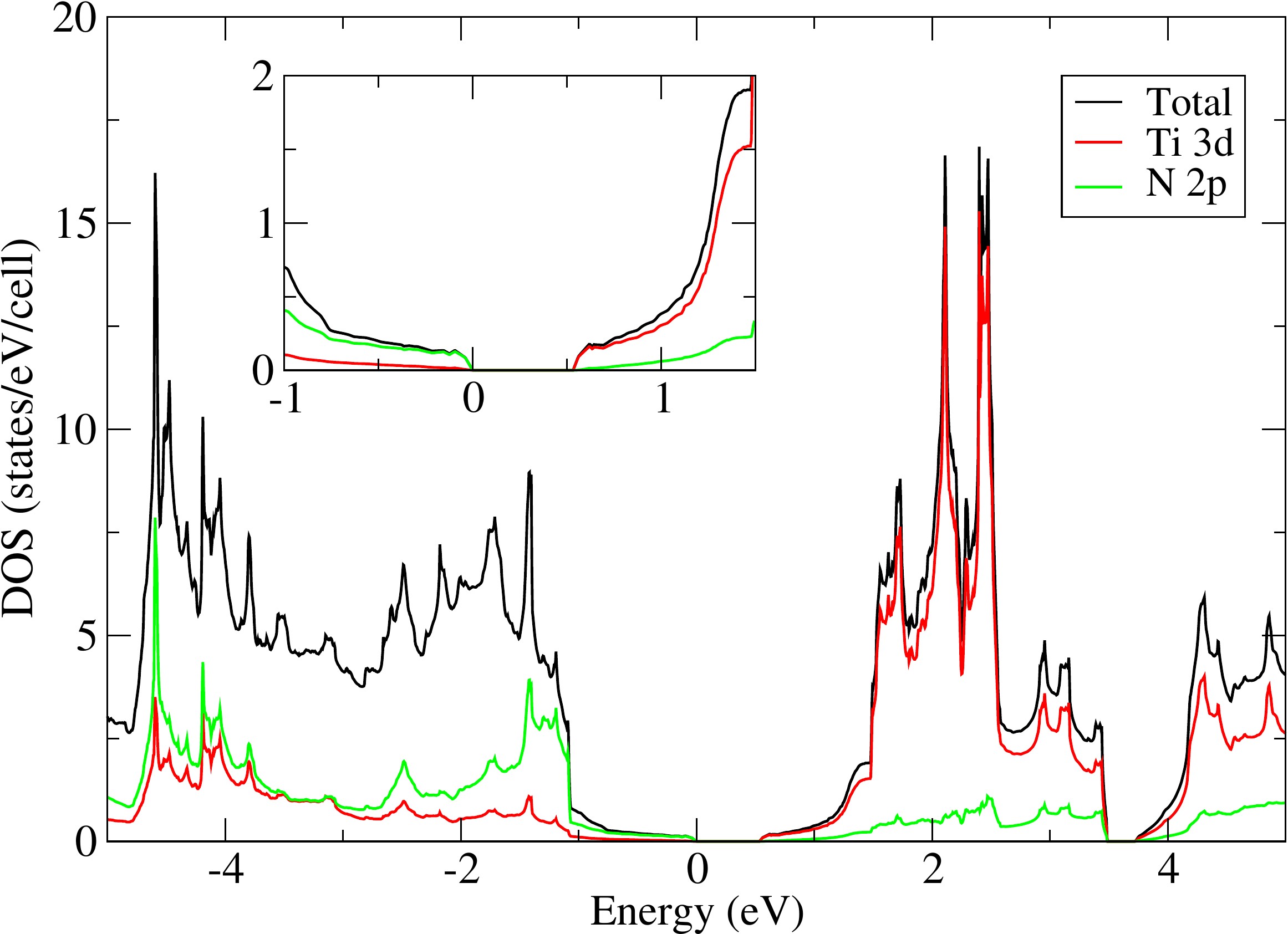}
\caption{(color online) Total and partial density of states of TiNCl, showing
the Ti $3d$ - N $2p$ mixing. The inset figure is a blown-up of the region near
Fermi level.}
\label{fig:DOS}
\end{figure}

The twelve valence bands are made of six N $2p$ and six Cl $3p$ states, and
the conduction band is comprised of ten Ti $3d$ states. The $3d$ bands show
a ``$t_{2g}-e_g$'' crystal field splitting (three states below and two above),
that arises in spite of the
nonequivalence of the five $3d$ orbitals in this structure. As can be seen in
the partial density of states plotted in Fig. \ref{fig:DOS}, there is $3d$
weight in the valence bands and N weight in the conduction bands, reflecting
substantial N $2p$ - Ti $3d$ hybridization in addition to the ionic
character reflected in their formal charges.

Whereas the $3d$ environment appears locally to be pseudo-cubic, the low
site symmetry severely splits the N $2p$ states, with $2p_x$ and $2p_y$ becoming
quite distinct. The top valence band is
primarily N $2p_{x}$ character, which extends down to $-5$ eV. The N $p_y$
and $p_z$ bands have their maximum 1 eV lower, and the Cl $2p$ weight is
concentrated at the bottom of the valence bands.

The inset in Fig. \ref{fig:DOS} shows an enlargement of the total and
atom-projected DOS around the Fermi energy. The onset at 0.5 eV and
the smooth slope to 1.2 eV is characteristic of a two-dimensional band which
becomes non-parabolic away from the band edge, and strongly so in the
1.2-1.5 eV region. At 1.5 eV the onset of the second band, with its much
heavier mass, is clear. However, the DOS does not have the sharp step at the 
top of the valence band that is characteristic of a 2D system.

The Ti $3d$ orbitals are lifted in degeneracy entirely by the orthorhombic
point group site symmetry, but as mentioned above the conduction bands are
separated by a crystal field analogous to cubic $t_{2g}-e_g$ splitting.
Checking the band character reveals that, in terms of $3d$ orbitals
expressed in terms of the orthorhombic coordinate axes, $%
d_{xy},d_{z^{2}},d_{xz}$ have most of the weight in the $0.5-3$ eV region,
and $d_{yz},d_{x^{2}-y^{2}}$ are in a higher energy window of $4-6$ eV. Thus
it is feasible, in a low-energy tight-binding model, to include only Ti $%
d_{xy},d_{z^{2}},d_{xz}$ and N $2p_{x}$ states. Plotted on top of the DFT
bands in Fig. \ref{fig:band}\ is the tight-binding fit using the Wannier
functions. The representation of the full $t_{2g}$ complex is excellent, as
is that of the top of the upper valence band.

The distance between TiNCl slabs and the weak inter-layer coupling allows
intercalation of alkali atoms, which act as electron donors. This
feature validates the rigid band shift approximation in simulating doping.
Doped-in electronic carriers will go into the single lowest conduction band,
which is quite two-dimensional as mentioned above but is dispersive within
the plane. This band has strong Ti $3d_{xy}$ character, similar to the
in-plane $4d$ character in ZrNCl. The Fermi surface of electron-doped TiNCl
is an oval centered at the $\Gamma $ point. This point has some relevance for the
superconductivity, since with a single Fermi surface there can be no nesting
of disconnected Fermi surfaces, such as are proposed\cite{Kuroki-FLEX}
to play an important role in many other layered superconductors, such as Fe pnictides, as well as 
$\beta $-structured ZrNCl and HfNCl. The similar characters of TiNCl and
ZrNCl, and their similar values of T$_{c}$, suggest that possible nesting of
Fermi surfaces is not an important feature for pairing in the materials.

\subsection{Wannier Functions}

Because the susceptibilities we will calculate have a number
of local orbital matrix elements equal to the 4th power of the
number of orbitals retained, we have calculated selected low-energy Wannier functions (WFs)
that will be used to construct our many-body Hamiltonian, using projections of the Bloch states
onto the corresponding atomic orbitals. The four atomic orbitals
mentioned above allow us to reproduce the bands on either side of the
gap: Ti $d_{xy},d_{z^{2}},d_{xz}$ and N $p_{x}$.
While the Ti
\textquotedblleft $t_{2g}$\textquotedblright\ orbitals are not optimal in
diagonalizing the local \textquotedblleft octahedral\textquotedblright\
symmetry, they and their relation to the $2p_{x}$ orbital are more readily
visualized. Since the RPA calculations described below were performed in the
electron-doped region where the Fermi level is shifted into the conduction
bands, considering only the N $p_x$ WF in the valence band is sufficient to 
understand the $\mathbf{q}$-dependence. 
Aside from being farther removed in energy, the remainder of
the valence bands form a complex of bands spread uniformly over the zone,
contributing little to any $\mathbf{q}$-dependence.

The Ti-N layer is strongly buckled and there are 2 Ti and
2 N atoms per unit cell with different $z$ coordinates. the actual
tight-binding model contains 8 bands and 8 WFs, but WFs on symmetry related
ions are symmetry equivalent. Overall the Wannier orbitals generate a well
represented band structure compared to the DFT bands within the energy
window of interest, as shown in Fig. \ref{fig:band}.

The hopping amplitudes of the Wannier orbitals are listed in Table \ref%
{hopping}. Hopping integrals smaller than $0.05$ eV were not listed because
they only marginally alter the band structure and obfuscate interpretation.
The on-site energies of the \textquotedblleft $t_{2g}$\textquotedblright\
orbitals are $2.25\pm 0.08$eV, lying within the largest peak of the DOS. The $p_{x}
$ energy is $-3.22$eV, in the middle of the valence bands. Thus there is a $%
5.5$eV separation of valence and conduction band centers, and a gap of $0.5$%
eV.

The dispersion within the pair of valence $p_{x}$ bands is represented
largely by hopping between neighboring $p_{x}$ WFs (recall, the $p_{x}$ WF
contains some $d$ character, and vice versa), both being about $0.5$eV.
Hopping amplitudes to the $d$ orbitals are $t_{pd}\approx 0.2-0.3$eV. In the
conduction bands, the $d_{xy}$ orbital has hopping amplitude $|t|\sim 0.17$%
eV to its partner within the cell as well as to its replicas in neighboring
cells in both directions. The hopping to the $p_{x}$ orbital (0.33 eV) is
twice as large, and apparently is the dominant contributor to the $3.5$eV
bandwidth. Due to the relative orientations, hopping to the other $d$ WFs is
no more than half as large as the $d_{xy}-d_{xy}$ one. The other two $d$ WFs
form rather narrow bands, reflected by smaller hopping amplitudes; note that
both have hopping to the $p_{x}$ orbital of $0.21-0.24$eV.

\begin{table}[tbh] \centering%
\begin{tabular}{l|cccccc}
\hline\hline
$(\mu ,\nu )$ & $[0,0,0]$ & $[1,0,\Delta z]$ & $[0,1,\Delta z]$ & $%
[1,1,\Delta z]$ & $[2,0,0]$ & $[0,2,0]$ \\ \hline
$(d_{xy},d_{xy})$ & \multicolumn{1}{|r}{$2.33$} & \multicolumn{1}{r}{} & 
\multicolumn{1}{r}{} & \multicolumn{1}{r}{$-0.16$} & \multicolumn{1}{r}{$%
-0.17$} & \multicolumn{1}{r}{$-0.18$} \\ 
$(d_{xy},d_{z^{2}})$ & \multicolumn{1}{|r}{} & \multicolumn{1}{r}{} & 
\multicolumn{1}{r}{} & \multicolumn{1}{r}{$-0.08$} & \multicolumn{1}{r}{} & 
\multicolumn{1}{r}{} \\ 
$(d_{xy},d_{xz})$ & \multicolumn{1}{|r}{} & \multicolumn{1}{r}{} & 
\multicolumn{1}{r}{} & \multicolumn{1}{r}{$0.07$} & \multicolumn{1}{r}{} & 
\multicolumn{1}{r}{$-0.11$} \\ 
$(d_{z^{2}},d_{z^{2}})$ & \multicolumn{1}{|r}{$2.18$} & \multicolumn{1}{r}{}
& \multicolumn{1}{r}{} & \multicolumn{1}{r}{$-0.05$} & \multicolumn{1}{r}{}
& \multicolumn{1}{r}{} \\ 
$(d_{z^{2}},d_{xz})$ & \multicolumn{1}{|r}{} & \multicolumn{1}{r}{} & 
\multicolumn{1}{r}{} & \multicolumn{1}{r}{$0.19$} & \multicolumn{1}{r}{$0.07$%
} & \multicolumn{1}{r}{} \\ 
$(d_{xz},d_{xz})$ & \multicolumn{1}{|r}{$2.25$} & \multicolumn{1}{r}{} & 
\multicolumn{1}{r}{} & \multicolumn{1}{r}{} & \multicolumn{1}{r}{$-0.05$} & 
\multicolumn{1}{r}{$0.13$} \\ \hline
$(d_{xy},p_{x})$ & \multicolumn{1}{|r}{} & \multicolumn{1}{r}{} & 
\multicolumn{1}{r}{$-0.33$} & \multicolumn{1}{r}{} & \multicolumn{1}{r}{} & 
\multicolumn{1}{r}{} \\ 
$(d_{z^{2}},p_{x})$ & \multicolumn{1}{|r}{} & \multicolumn{1}{r}{$0.24$} & 
\multicolumn{1}{r}{$0.22$} & \multicolumn{1}{r}{} & \multicolumn{1}{r}{} & 
\multicolumn{1}{r}{} \\ 
$(d_{xz},p_{x})$ & \multicolumn{1}{|r}{} & \multicolumn{1}{r}{} & 
\multicolumn{1}{r}{$0.21$} & \multicolumn{1}{r}{} & \multicolumn{1}{r}{} & 
\multicolumn{1}{r}{} \\ \hline
$(p_{x},p_{x})$ & \multicolumn{1}{|r}{$-3.22$} & \multicolumn{1}{r}{} & 
\multicolumn{1}{r}{} & \multicolumn{1}{r}{$0.51$} & \multicolumn{1}{r}{$0.47$%
} & \multicolumn{1}{r}{$0.14$} \\ \hline\hline
\end{tabular}%
\caption{Nearest, 2nd, and 3rd neighbor hopping integrals in units of eV.
The hopping vectors are in units of [a/2, b/2, c], and $\Delta$z represents
the difference between the z coordinates of the two orbitals. This representation
is purely two-dimension (no coupling along the $c$ axis).}\label{hopping}%
\end{table}%

\section{Many-body Hamiltonian and Random Phase Approximation}

The random phase approximation (RPA) applies an interaction to the
non-interacting Hamiltonian%
\begin{equation}
H_{0}=\sum_{\mathbf{k},ab}H_{ab}^{\mathbf{k}}c_{\mathbf{k},a}^{\dagger} c_{%
\mathbf{k},b}^{},  \label{H0}
\end{equation}%
where $a,b$ are composite orbital and spin indices of the basis Wannier
orbitals. The interaction Hamiltonian, in general, is the following
symmetric form%
\begin{equation}
H_{1}=\frac{1}{2}\sum_{i}\sum_{abcd}U_{abcd}c_{ia}^{\dagger}c_{ib}^{}
c_{ic}^{\dagger}c_{id}^{}+\sum_{\langle i,j \rangle}\sum_{abcd}V_{abcd}c_{ia}^{%
\dagger}c_{ib}^{} c_{jc}^{\dagger}c_{jd}^{},  \label{H1}
\end{equation}%
where $U$ and $V$ represent on-site and inter-site (only nearest neighbors)
interactions, respectively. (\ref{H1}) can be Fourier transformed into%
\begin{equation}
H_{1}=\frac{1}{2N}\sum_{\mathbf{kpq}}\sum_{abcd}F_{abcd}(\mathbf{q})c_{%
\mathbf{k},a}^{\dagger}c_{\mathbf{k+q},b}^{}c_{\mathbf{p+q,}c}^{\dagger}c_{%
\mathbf{p}d}^{},  \label{H1(k)}
\end{equation}%
in which $F_{abcd}(\mathbf{q})=U_{abcd}+\gamma (\mathbf{q})V_{abcd}$ is the
interaction kernel matrix. In the second term, $\gamma (\mathbf{q}%
)=\sum_{l}e^{i\mathbf{q}\cdot \mathbf{R}_{l}}$ ($l$ running over the nearest
neighbor pairs of sites) is the structure factor, which brings in
$\mathbf{q}$-dependence into $F_{abcd}(\mathbf{q})$. The bare susceptibility is
calculated as%
\begin{equation}
\chi _{abcd}^{0}(\mathbf{q},\omega)=\sum_{\mathbf{k}}G_{ad}(\mathbf{k}%
,\omega )G_{cb}(\mathbf{k+q},\omega ),  \label{X0}
\end{equation}%
where $G_{ab}(\mathbf{k},\omega )$ is the non-interacting Green's function%
\begin{equation}
G_{ab}(\mathbf{k},\omega )=\sum_{n}\frac{\left\langle a\right\vert n\mathbf{k%
}\rangle \langle n\mathbf{k}\left\vert b\right\rangle }{\omega +\mu
-\varepsilon _{n\mathbf{k}}},  \label{G0}
\end{equation}%
and the summation is taken over all bands. Applying RPA, which sums up the
higher order diagrams in the geometric series, we have the full
susceptibility presented in a matrix equation%
\begin{equation}
\chi(\mathbf{q}, \omega )= [I+\chi^{0}(\mathbf{q},\omega)\,\mathrm{Re}\,F(%
\mathbf{q})]^{-1}\chi^{0}(\mathbf{q},\omega)  \label{X}
\end{equation}
where a matrix $\chi$ is formed from $\chi_{abcd}$ by contracting the first
pair of indices and the last pair of indices.

So far we have derived a very general formula for an arbitrary interaction
Hamiltonian. To study our case, next consider a more specific model in the
form of an extended Hubbard Hamiltonian, following Kuroki's model \cite%
{Kuroki-FeAs}, but add an extra inter-site interaction term:%
\begin{eqnarray}
H_{1} &=&\sum_{i} \left[ U\sum_{a}n_{ia\uparrow }n_{ia\downarrow }+U^{\prime
}\sum_{a\neq b}\sum_{\sigma ,\sigma ^{\prime }}n_{ia\sigma }n_{ib\sigma
^{\prime }}  \label{H1ex} \right. \\
&& \left. -J\sum_{a\neq b}S_{ia}\cdot S_{ib}+J^{\prime }\sum_{a\neq b}c_{ia\uparrow
}^{\dagger }c_{ia\downarrow }^{\dagger }c_{ib\downarrow }^{{}}c_{ib\uparrow
}^{{}} \right] \notag \\
&&+\sum_{\langle i,j \rangle}\sum_{a,b}V_{ab}n_{ia}n_{jb},  \notag
\end{eqnarray}%
in which $a$, $b$ are orbital indices, $i$, $j$ are site indices of the
lattice, and $\sigma $ is the spin index. $U$ is the intra-orbital Coulomb
repulsion, $U^{\prime }$ is the inter-orbital Coulomb interaction, $%
t_{ij}^{\mu \nu }$ is the hopping between Wannier orbitals, $V_{ab}$ is the
inter-site Coulomb interaction between orbitals $a$ and $b$, $J$ is the
Hund's rule coupling, and $J^{\prime }$ is referred to pair hopping between
orbitals. From this Hamiltonian, the susceptibility is calculated by%
\begin{eqnarray}
\chi ^{S}(\omega ,\mathbf{q}) &=&\frac{\chi ^{0}(\omega ,\mathbf{q})}{I-S(%
\mathbf{q})\chi ^{0}(\omega ,\mathbf{q})},  \label{XsXc} \\
\chi ^{C}(\omega ,\mathbf{q}) &=&\frac{\chi ^{0}(\omega ,\mathbf{q})}{I+C(%
\mathbf{q})\chi ^{0}(\omega ,\mathbf{q})}.  \notag
\end{eqnarray}%
The interaction matrices $S(\mathbf{q})$ and $C(\mathbf{q})$ take the form:%
\begin{eqnarray}
S_{abcd} &=&\left\{ 
\begin{array}{cc}
U, & a=b=c=d \\ 
U^{\prime }, & a=c\neq b=d \\ 
J, & a=b\neq c=d \\ 
J^{\prime }, & a=d\neq b=c%
\end{array}%
\right\} ,  \label{S-C} \\
C_{abcd} &=&\left\{ 
\begin{array}{cc}
U+2V_{ac}\,\mathrm{Re}\,\gamma (\mathbf{q}), & a=b=c=d \\ 
-U^{\prime }+J, & a=c\neq b=d \\ 
2U^{\prime }-J+2V_{ac}\,\mathrm{Re}\,\gamma (\mathbf{q}), & a=b\neq c=d \\ 
J^{\prime }, & a=d\neq b=c%
\end{array}%
\right\} ,  \notag
\end{eqnarray}%
where $U$, $U^{\prime }$, $J$, and $J^{\prime }$ terms appear only if all
indices are orbitals on the same site. Finally, we can also calculate the
macroscopic susceptibilties by performing a summation over the orbital
indexes:%
\begin{equation}
\chi ^{mac}(\mathbf{q},\omega )=\sum_{ijkl}S_{ij}\chi _{ijkl}(\mathbf{q}%
,\omega )S_{kl},  \label{X-mac}
\end{equation}%
in which $S_{ij}$ is overlap matrix, and in our case it has the form of
delta-function $\delta _{ij}$.

\section{Spin and Charge Susceptibility}

With our model just constructed, we calculate the spin and charge
susceptibilities. The model is a multi-band extended Hubbard model on a 2D
rectangular lattice with 4 sites (two Ti and two N) per unit cell. For the on-site interaction
terms, we use $U_{dd}=U_{dd}^{\prime }=1.5$ eV, $U_{pp}=1.0$ eV, and $%
J=J^{\prime }=0.2$ eV. These values are somewhat smaller than might be used
in a traditional Hubbard model calculation; this is partly to compensate for
the fact that RPA has a tendency to overestimate the strength of the
interaction due to the lack of the self-energy correction.\cite{Kuroki-FeAs}
Moreover, we are using WFs rather than atomic orbitals, for which the
extension onto neighboring sites will suppress the intra-atomic interactions $%
U$ and $J$.

For inter-site interactions, we assume $V_{ac}$ to be spin and orbital
independent, and that it only depends on the distance between the two sites.
Taking into account the Ti-N and Ti-Ti distances mentioned above, we use $%
V_{Ti-N}=0.5$ eV (nearest neighbor), $V_{Ti-Ti}=0.3$ eV (2nd nearest
neighbor). Since the Wannier functions have contributions from neighboring
sites, it is reasonable to set the inter-site Coulomb repulsion $V$ slightly
larger than traditionally used for atomic orbitals. The calculation is done
at $T=0.02$ eV ($220K$) and $\omega =0$, with a \textbf{k}-mesh of $40\times
40\times 4$ and \textbf{q}-mesh of $20\times 20\times 2.$ The occupation is set at $%
4.3$, simulating $x=0.15$ electron doping in $A_{0.15}$TiNCl (since there
are two formula units per unit cell) by raising 
the Fermi level into the lowest conduction band.

\begin{figure}[b]
\centering
\includegraphics[angle=-90,width=0.6\columnwidth]{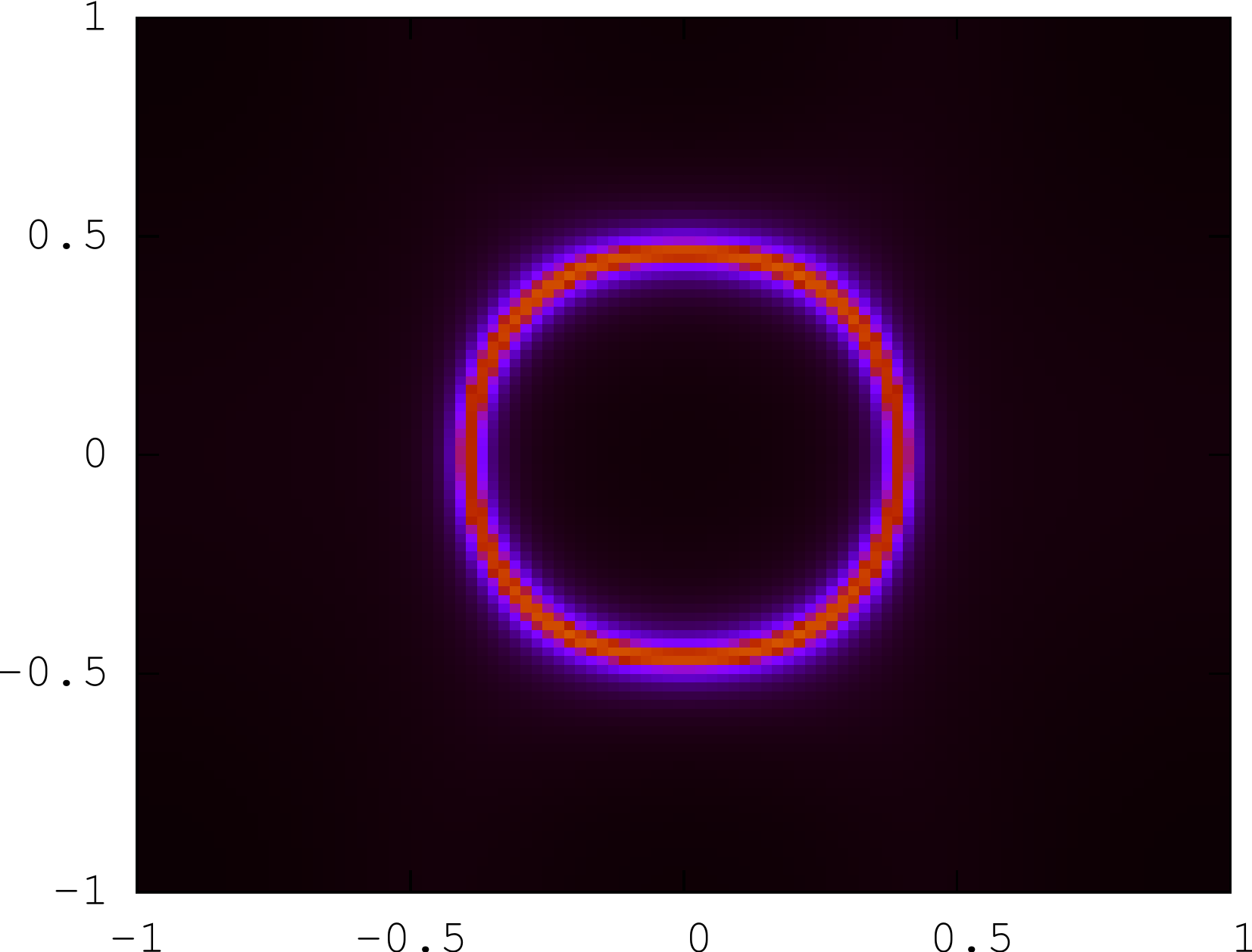}
\caption{(color online) The spectral function $-\mathrm{Im}\, G(\mathbf{k},\protect\omega =0)$
in the $k_x-k_y$ plane, showing the oval (nearly circular) Fermi surface for $x$=0.15 electron doping.}
\label{fig:fermisf}
\end{figure}

Figure \ref{fig:fermisf} shows the
magnitude of the imaginary part of Green's function, which provides a view
of the Fermi surface. With a simple, nearly circular Fermi surface like
this, the bare susceptibility is expected\cite{2D} to be isotropic out to $%
q=2k_{F}$, with a relatively constant plateau behavior inside $2k_{F}$
radius. The inter-site Coulomb interaction can give rise to charge
fluctuation, creating collective electron motion and possible charge ordering.
Competition between on-site and inter-site interaction of $d$ electrons can
lead in principle to a frustration of both spin and charge ordering. 
The hybridization between $d$ and $p$
orbitals opens another channel, that of a charge transfer instability. One
of the interesting questions is whether some combination of these processes can create
excitations that can pair up electrons, analogous to the behavior found in a $d$-$%
p$ model in the limit of infinite $U$ and nearest neighbor hybridization 
\cite{ChargeFluc3}.

\begin{figure}[tbh]
\begin{center}
\subfigure[]{\includegraphics[angle=-90,width=0.75\columnwidth,draft=false]{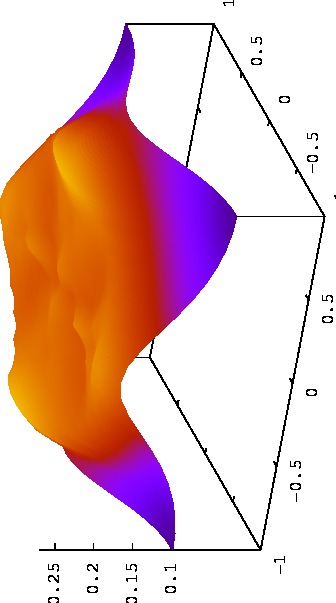}}
\subfigure[]{\includegraphics[angle=-90,width=0.75\columnwidth,draft=false]{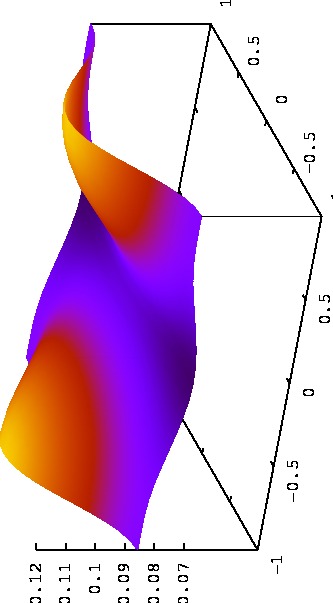}}
\subfigure[]{\includegraphics[angle=-90,width=0.75\columnwidth,draft=false]{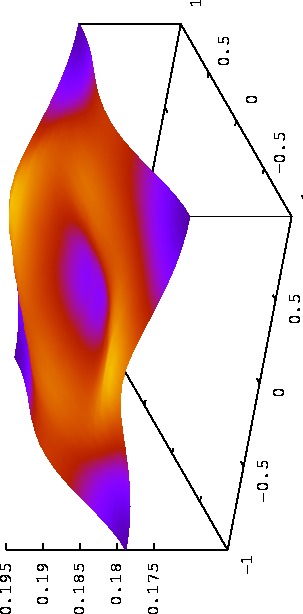}}
\end{center}
\caption{(color online) Representative orbital spin susceptibilities. In the basal plane, the
axis on the left (longer in perspective) is $q_{x}$ axis, and the one on the right (shorter) is $q_{y}$ axis.
(a) $\protect\chi _{1111}^{S}$ (b) $\protect\chi _{1133}^{S}$ (c) $\protect\chi _{1177}^{S}$.
Note that  the maximum in $\protect\chi _{1133}^{S}$ is located at $(\pi/a,0,0)$ and not at $2k_{F}$.
Note the different vertical scales on the panels.}
\label{fig:Xs}
\end{figure}

In Figure \ref{fig:Xs} some representative spin susceptibilities in the orbital
representation are plotted on the two-dimensional basal plane ($q_{z}=0$) in
the BZ. To clarify, we denote the orbitals by numbers in the order: (1)Ti$%
_{1}$-$d_{xy}$, (2)Ti$_{2}$-$d_{xy}$, (3)Ti$_{1}$-$d_{z^{2}}$, (4)Ti$_{2}$-$%
d_{z^{2}}$, (5)Ti$_{1}$-$d_{xz}$, (6)Ti$_{2}$-$d_{xz}$, (7)N$_{1}$-$p_{x}$,
(8)N$_{2}$-$p_{x}$. The largest spin susceptibility is found for $\chi _{1111}^{S}$
(intersite, $xy \leftrightarrow xy$) which has approximate 4-fold 
symmetry for magnetic fluctuations of 
the same orbital $d_{xy}$ on Ti sites.
The anisotropic behavior of $\chi_{1133}^{S}$ (on-site, $xy \leftrightarrow z^2$)
comes from the orthorhombic symmetry
of the lattice, i.e. $a\neq b$, which brings in anisotropic $\mathbf{q}$-dependence, in
this case strongly so.
The spin fluctuations between $d$ and $p$ orbitals $\chi _{1177}^{S}$ have a sizable
overall magnitude, comparable to $d-d$ fluctuation, but small variation 
with $\mathbf{q}$, because four
neighboring N atoms have almost the same distance to Ti. Due to the lack of
Fermi surface nesting, there is no divergent behavior in the spin
susceptibility, presenting different physics from the $\beta $-HfNCl which
has two circular Fermi surfaces located at two high-symmetry (K) points in
the BZ which can provide near perfect nesting.

\begin{figure}[tbh]
\begin{center}
\subfigure[]{\includegraphics[angle=-90,width=0.75\columnwidth,draft=false]{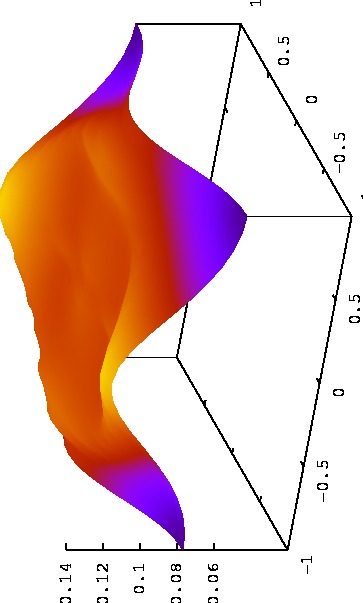}}
\subfigure[]{\includegraphics[angle=-90,width=0.75\columnwidth,draft=false]{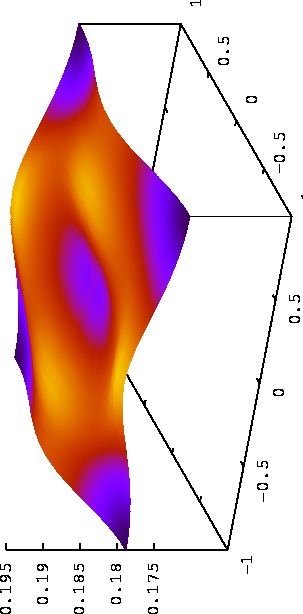}}
\end{center}
\caption{(color online) Representative charge susceptibilities in the full Brillouin zone.
(a) $\protect\chi _{1111}^{C}$
(b) $\protect\chi _{1177}^{C}$. The latter quantity is hardly distinguishable from its
spin counterpart.  Note the different vertical scales on the panels.}
\label{fig:Xc}
\end{figure}

Representative charge susceptibilities are shown in Figure \ref{fig:Xc}. Often
they have similiarities to the spin susceptibilities, with comparable 
but somewhat smaller
magnitudes. It is known that in an extended Hubbard
model on a square lattice, at zero frequency, $\chi^{S}$ and $\chi^{C}$
have similar $\mathbf{q}$-dependence and only vary in magnitude.\cite{Charge-fluc-d} Without the
long-range Coulomb interaction, charge fluctuation will always be 
smaller than spin fluctuation because of the different signs in the RPA
formula. The difference between $\chi^{S}$ and $\chi^{C}$ will become
more apparent at non-zero $\omega $. The $\mathbf{q}$-dependence of $\chi^{C}$ is
similar to that of $\chi^{S}$ but shows somewhat more structure in $\chi _{1111}^{C}$.
Note that the magnitude of  $\chi _{1111}^{C}$ is only half that of $\chi _{1111}^{S}$.
The spin and charge fluctuations within the N $p$%
-channel are very small since the $p$ orbitals are fully occupied so 
fluctuations only happen as a second order effect. However, the presence of
the N $p$ band very close to the lowest $d$ conduction band opens an
additional channel for fluctuations between them ($\chi_{1177}^S$, Fig. \ref{fig:Xs}c and $\chi_{1177}^{C}$, Fig. \ref{fig:Xc}b).

\begin{figure}[t]
\begin{center}
\subfigure[]{\includegraphics[angle=-90,width=0.75\columnwidth,draft=false]{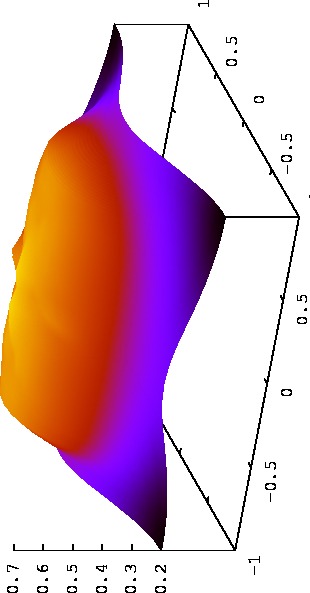}}
\subfigure[]{\includegraphics[angle=-90,width=0.75\columnwidth,draft=false]{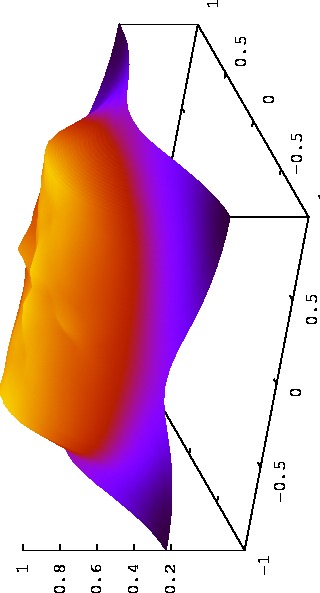}}
\subfigure[]{\includegraphics[angle=-90,width=0.75\columnwidth,draft=false]{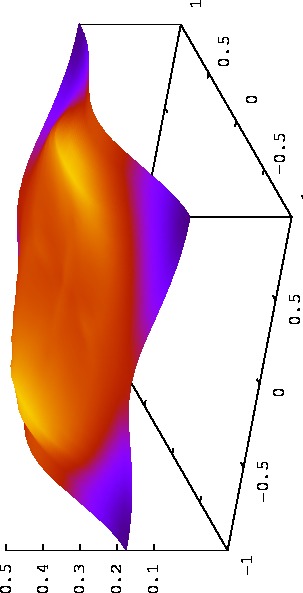}}
\end{center}
\caption{(color online) Macroscopic susceptibility: (a) bare; (b) spin; (c) charge.  Note the strong deviation from square symmetry.}
\label{fig:Xmac}
\end{figure}

To close the comparison, we show the macroscopic susceptibilities in Figure \ref{fig:Xmac}. The imprint of $2k_{F}$ is evident.
Beyond $2k_{F}$, the susceptibilities decrease slightly and with (near) square symmetry.
Inside $2k_{F}$, the variation is greater and displays the rectangular symmetry of the lattice. The
overall spin enhancement of the macroscopic susceptibility ($\chi^{S}(q)/\chi^{0}(q)$) near $\mathbf{q} = 0$ is
about $1.4$. For the model with the realistic parameters considered here, our test calculations
show that $\chi^{S}$ will approach divergent behavior when $U \sim 4$ eV. Since TiNCl has wide
$3d$ bands, it is unphysical to use $U$ anywhere near $4$ eV (we are using 1.5 eV), so strong spin fluctuation 
is unlikely to occur in this material. The macroscopic charge susceptibility has smaller magnitude
and $\mathbf{q}$-dependence, but will show divergent behavior when inter-site interaction $V$
is much larger than $U$, but that regime is unrealistic for the case studied here. Overall, both spin
and charge susceptibilities show moderate enhancements compared to the bare susceptibility, without
any approach to an instability toward spin or charge ordering.

\section{Conclusion}

To conclude, we have constructed a many-body extended Hubbard model using a realistic
band structure obtained from density functional theory calculations. RPA is
applied to obtain the spin and charge susceptibilities. In a system like $%
\alpha $-TiNCl, where the crucial ingredients of high-temperature
superconductivity, such as strong electron-phonon coupling and good Fermi
surface nesting, seem to be missing, spin and charge fluctuations are
the remaining candidates. Our calculations show that the spin and charge
enhancements of susceptibilities, both intra-band and inter-band, are 
small due to moderate correlations.
However, spin and charge fluctuations can 
produce substantial values possibly capable of encouraging electrons to pair.
Although spin fluctuations are present in $\alpha $%
-TiNCl, it is worth pointing out that the physics is very different from the 
$\beta $-structure counterparts even though both are nonmagnetic, and
apparently distinct from the recently discovered Fe pnictides where
magnetism is an important feature in parent compounds. As the nonmagnetic
nature of TiNCl and other MNX materials indicates, as well as seen from the
results from our RPA calculation, charge fluctuations may have an important
role in  superconductivity in these systems. Overall, we
still do not have a clear understanding of how superconductivity arises from
the fluctuations, as with all other high-Tc families.

\section{Acknowledgment}

The authors would like to thank K. Kuroki and R. T. Scalettar for helpful
discussions on the technical details of the RPA calculations. This work is
supported by DOE Grants DE-FG02-04ER46111 and DE-FC02-06ER25794.


\begin{thebibliography}{99}
\bibitem{HfNCl-band} R. Weht, A. Filippetti, and W. E. Pickett, Europhys.
Lett. \textbf{48}, 320 (1999).

\bibitem{HfNCl-Nature1998} S. Yamanaka, K. Hotehama, and H. Kawaji, Nature
London \textbf{392}, 580 (1998).

\bibitem{beta-structure} X. Chen, T. Koiwasaki. and S. Yamanaka, J. Phys.:
Condens. Matter \textbf{14}, 11209 (2002).

\bibitem{alpha-structure} Shoji Yamanaka, Kojiro Itoh, Hiroshi Fukuoka, and
Masahiro Yasukawa, Inorg. Chem. \textbf{39}, 806 (2000).

\bibitem{LixZrNCl-Tcdoping} Y. Taguchi, A. Kitora, and Y. Iwasa, Phys. Rev.
Lett. \textbf{97}, 107001 (2006).

\bibitem{HfNCl-isotope} Hideki Tou, Yutaka Maniwa, and Shoji Yamanaka, Phys.
Rev. B \textbf{67}, 100509(R) (2003).

\bibitem{LixZrNCl-phonon} R. Heid and K.-P. Bohnen, Phys. Rev. B \textbf{72}%
, 134527 (2005).

\bibitem{TiNCl-SC} Shoji Yamanaka, Toshihiro Yasunaga, Kosuke Yamaguchi and
Masahiro Tagawa, J. Mater. Chem. \textbf{19}, 2573 (2009).

\bibitem{LixZrNCl-Xs-expt} Yuichi Kasahara, Tsukasa Kishiume, Takumi Takano,
Katsuki Kobayashi, Eiichi Matsuoka, Hideya Onodera, Kazuhiko Kuroki,
Yasujiro Taguchi, and Yoshihiro Iwasa, Phys. Rev. Lett. \textbf{103}, 077004
(2009).

\bibitem{HfNCl-Tcspacing} T. Takano, T. Kishiume, Y. Taguchi, and Y. Iwasa,
Phys. Rev. Lett. \textbf{100}, 247005 (2008).

\bibitem{LixZrNCl-spheat} Y. Taguchi, M. Hisakabe, and Y. Iwasa, Phys. Rev.
Lett. \textbf{94}, 217002 (2005).

\bibitem{Kuroki-FLEX} K. Kuroki, Sci. Tech. Adv. Mater. \textbf{9}, 044202
(2008).

\bibitem{plasmon1} A. Bill, H. Morawitz, and V. Z. Kresin, Phys. Rev. B 
\textbf{66}, 100501(R) (2002).

\bibitem{plasmon2} A. Bill, H. Morawitz, and V. Z. Kresin, Phys. Rev. B 
\textbf{68}, 144519 (2003).

\bibitem{NaCoO2-FLEX} Masahito Mochizuki, Youichi Yanase, and Masao Ogata,
Phys. Rev. Lett. \textbf{94}, 147005 (2005).

\bibitem{NaCoO2-RPA} Yasuhiro Tanaka , Yoichi Yanase, Masao Ogata, Physica B 
\textbf{359}--\textbf{361}, 591 (2005).

\bibitem{ChargeFluc-organicSC} Jaime Merino and Ross H. McKenzie, Phys. Rev.
Lett. \textbf{87, }237002 (2001).

\bibitem{ChargeFluc1} F. Bucci, C. Castellani, C. Di Castro, and M. Grilli,
Phys. Rev. B \textbf{52}, 6880 (1995).

\bibitem{ChargeFluc2} Y. M. Vilk, Liang Chen, and A. M. S. Tremblay, Phys.
Rev. B \textbf{49}, 13267 (1994).

\bibitem{2D} W. E. Pickett, J. Supercond. \& Novel Magn. \textbf{19}, 291
(2006).

\bibitem{ChargeFluc3} Munehiro Azami, Akito Kobayashi, Tamifusa Matsuura,
Yoshihiro Kuroda, Physica C \textbf{259,} 227 (1996).

\bibitem{LixZrNCl-isotope} Y. Taguchi, T. Kawabata, T. Takano, A. Kitora, K.
Kato, M. Takata, and Y. Iwasa, Phys. Rev. B \textbf{76}, 064508 (2007).

\bibitem{Li(THF)HfNCl-X-expt} H. Tou, Y. Maniwa, T. Koiwasaki, and S.
Yamanaka, Phys. Rev. Lett. \textbf{86}, 5775 (2001).

\bibitem{LiZrNCl-SCgap} Tomoaki Takasaki, Toshikazu Ekino, Hironobu Fujii
and Shoji Yamanaka, J. Phys. Soc. Japan \textbf{74}, 2586 (2005).

\bibitem{LiTHFHfNCl-gap1} T. Ekino, T. Takasaki, H. Fujii, and S. Yamanaka,
Physica C \textbf{388-389}, 573 (2003).

\bibitem{LiTHFHfNCl-gap2} T. Ekinoa, T. Takasaki, T. Muranaka, H. Fujii, J.
Akimitsu, S. Yamanaka, Physica B \textbf{328}, 23 (2003).

\bibitem{Yamanaka-review} Shoji Yamanaka, Annu. Rev. Mater. Sci. \textbf{30,}
53 (2000).

\bibitem{FPLO} K. Koepernik and H. Eschrig, Phys. Rev. B \textbf{59}, 1743
(1999).

\bibitem{GGA96} J. P. Perdew, K. Burke, and M. Ernzerhof, Phys. Rev. Lett. 
\textbf{77}, 3865 (1996).

\bibitem{Kuroki-FeAs} Kazuhiko Kuroki, Hidetomo Usui, Seiichiro Onari,
Ryotaro Arita, and Hideo Aoki, Phys. Rev. B \textbf{79}, 224511 (2009).

\bibitem{ZrHfNCl-band1} Izumi Hase and Yoshikazu Nishihara, Phys. Rev. B 
\textbf{60}, 1573 (1999).

\bibitem{ZrHfNCl-band2} Izumi Hase and Yoshikazu Nishihara, Physica B 
\textbf{281\&282, }788 (2000).

\bibitem{MNCl1-x-SC} S. Yamanaka, L. Zhua, X. Chena, H. Tou, Physica B 
\textbf{6--9}, 328 (2003).

\bibitem{ZrNCl-band} Haruka Sugimoto and Tamio Oguchi, J. Phys. Soc. Jpn. 
\textbf{73}, 2771 (2004).

\bibitem{Charge-fluc-d} Khee-Kyun Voo and W. C. Wu, Jian-Xin Li and T. K.
Lee, Phys. Rev. B \textbf{61}, 9095 (2000).

\bibitem{Jain1985} J. K. Jain and P. B. Allen, 
Phys. Rev. B \textbf{32}, 997 (1985)

\end{thebibliography}
\end{document}